\documentclass[%
 aps,
 amsmath,amssymb,
preprint,%
]{revtex4-1}

\usepackage{graphicx}
\usepackage{dcolumn}
\usepackage{bm}

\begin{document}


\title{Self Healing Interconnects with a Near Plastically Stretchable Heal}

\author{Amit Kumar}
\author{Virendra Parab}
\author{Arindan Handu}
\affiliation{Department of Instrumentation and Applied Physics, Indian Institute of Science, Bangalore, India}
\author{Li Ding}
\affiliation{Department of Engineering, University of Cambridge}
\author{Sanjiv Sambandan}
\affiliation{Department of Engineering, University of Cambridge}
\affiliation{Department of Instrumentation and Applied Physics, Indian Institute of Science, Bangalore, India}
\email{ss698@cam.ac.uk}

\date{\today}

\begin{abstract}
Flexible electronic systems such as roll up displays, wearable devices etc. promise exciting possibilities that could change the way humans interact with the environment. However, they suffer from poor reliability of interconnects and devices. Interconnects on flex are prone to open circuit failures due to mechanical stress, electrostatic discharge and environmental degradation. Passive approaches such as the use of stretchable conductors and novel geometries improve their response to mechanical stress but cannot salvage the interconnect if a fault were to occur. Active approaches using self healing techniques can repair a fault and have been demonstrated using methods that either use relatively rare materials, change conventional interconnect fabrication processes, only address faults due to mechanical stress or do not permit stretching. In this work we discuss a self healing technique that overcomes these limitations and demonstrate heals having metallic conductivity and nearly plastic stretchability. This is achieved using a dispersion of conductive particles in an insulating fluid encapsulated over the interconnect. Healing is automatically triggered by the electric field appearing in the open gap of a failed interconnect, irrespective of the cause of failure. The field polarizes the conductive particles causing them aggregate and chain up to bridge the gap and repair the fault. Using copper-silicone oil dispersions, we show self healing interconnects with the stretchable heal having conductivity of about $5 \times 10^{5}$ Sm and allowing strains from 12 to 60. Previously, stretchable interconnects used materials other than copper. Here we effectively show self healing, stretchable copper. This work promises high speed, self healing and stretchable interconnects on flex thereby improving system reliability.
\end{abstract}

\maketitle

\section{Introduction}
Electronics on flex offer aesthetic and functional advantages such as the possibility of roll to roll manufacturing \cite{roll2}, use of organic metals and semiconductors \cite{organic1, organic2, flex1}, sensors and actuators on flex \cite{flex1, flex4, flex5, flex8}, energy sources \cite{energy1, energy2, energy3} and wearable devices for health diagnostics \cite{ wear2, wear3, wear4, wear5}. However, these advantages are accompanied by poor reliability of devices \cite{defect1} and interconnects \cite{SanjivDefectIdentification}. Interconnects on flex experience mechanical forces due to stretching, bending and thermal stress; unexpected current surges due to electrostatic discharge and environment related degradation. As a result, open interconnect faults occurring during system operation are common.

Both passive and active techniques have been investigated to address this problem. Passive approaches tailor interconnect geometries \cite{flex1}, \cite{geometry0, geometry1, geometry3, geometry4, geometry5, geometry6, geometry7, geometry8, geometry9} and materials \cite{geometry3},\cite{stretch1, stretch2, stretch4, stretch5, stretch8, stretch9, stretch11, stretch12, stretch13, stretch14,  stretch16, stretch17, stretch18, stretch19} to improve mechanical flexibility. While they do improve tolerance to mechanical stress driven faults, they do not prevent faults occurring due to other mechanisms. Moreover, they do not provide a means to re-establish connectivity if an open fault does occur. On the other hand, active techniques offer on-line repair of an open fault. Self healing of interconnects using conductive polymers \cite{selfhealGeneral1}, \cite{selfhealGeneral2}, liquid metals \cite{selfhealLM1, selfhealLM2, selfheal1}, embedding of capsules of conductive inks in interconnects that spill the ink upon fracture \cite{selfheal2, selfheal3, selfheal4}, controller based impedance modulation \cite{selfheal54}, Janus particles driven by differences in hydrophobicities \cite{selfheal6},  ionic gels \cite{selfheal5} and electric field driven interconnect re-structuring using dispersions have been investigated \cite{selfheal7, selfheal8, selfheal9, selfheal10}. While these approaches have been shown to be very effective, the techniques have some shortcomings as they either use relatively rare materials (eg. Ga, In), change conventional interconnect fabrication processes, only address mechanical stress related faults or do not permit stretchability. 

An ideal solution towards improving open fault tolerance of interconnects on flex would have the following attributes. First it would be a self driven repair mechanism. Second, it would permit a heal having near metallic conductivity. Third, it would permit a heal having high stretchability. Fourth, the repair mechanism would be activated irrespective of the cause of the fault.

\begin{figure*}
\centering
\includegraphics[width=6 in]{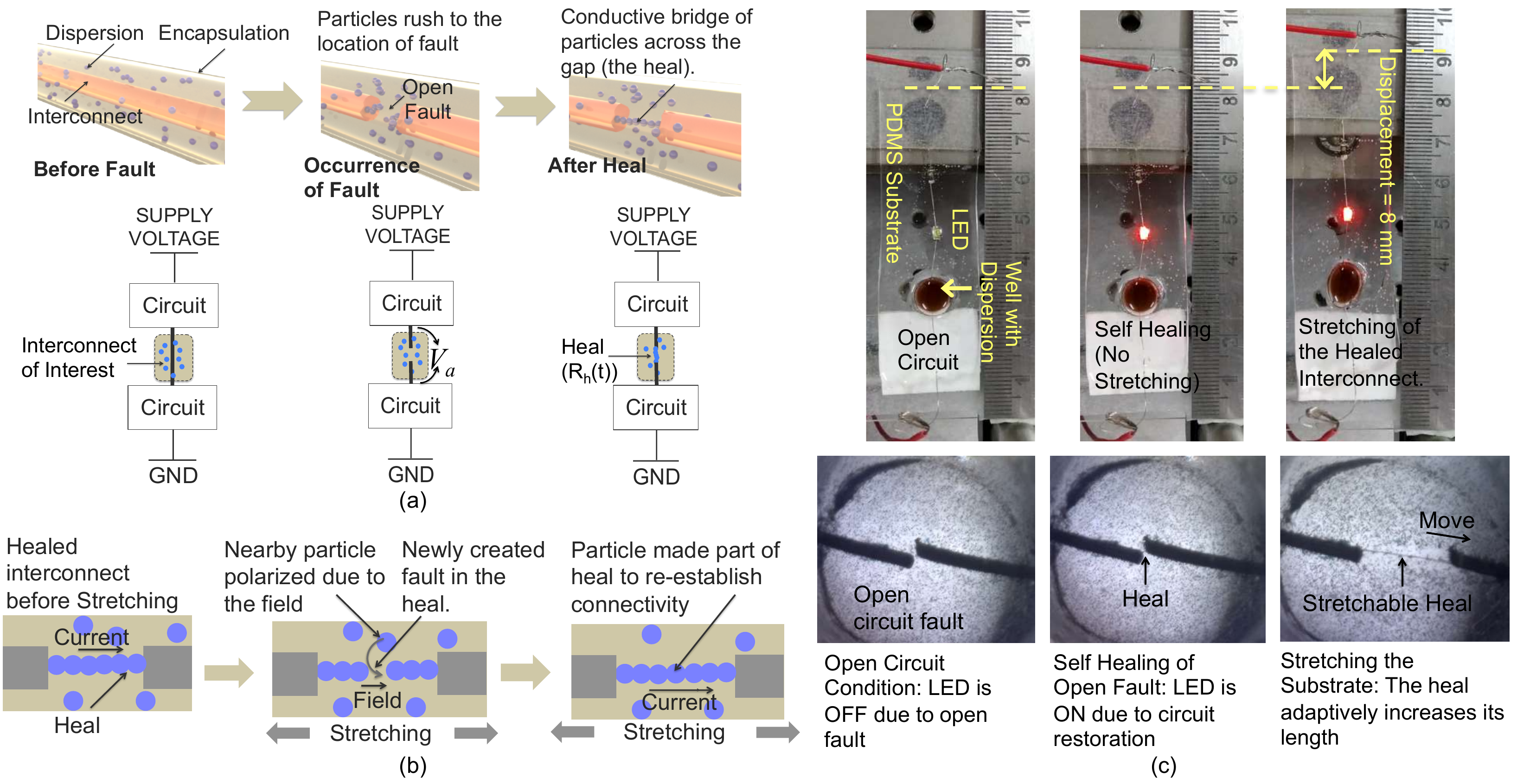}
\caption{(a) Illustration of the self healing mechanism. (b) Demonstration of stretchable self healing with the heal adaptively increasing in length to accommodate the stretching. The dispersion is contained in a well and over an interconnect with an artificially induced open fault. Initially, the fault is healed. Upon stretching the substrate, the heal stretches while maintaining electrical connectivity. The photographs of the events occurring inside the well are also shown. Videos in Supplementary Material.}
\end{figure*}

This work provides an affirmative answer to all four points. Here we use a dispersion of electrically conductive particles in an insulating fluid that is contained and isolated over each interconnect as shown in Fig. 1a. Upon the occurrence of an open circuit fault in a current carrying interconnect, the electric field appearing across the open gap polarizes the conductive particles of the dispersion inside the gap. The polarized particles experience dipole-dipole attractive forces and eventually chain up and sinter to create a bridge across the gap thereby healing the fault. Therefore the repair mechanism is independent of the cause the fault. This mechanism also permits near plastic stretchability (Fig. 1b). When the substrate is stretched, the heal momentarily breaks. This re-establishes the field in this newly formed break, polarizes locally available free partcles and encourages them to fill the gap to re-establish connectivity. The heal therefore stretches by increasing the number of particles constituting the chain thereby making this a unique approach to stretchable self healing. Fig. 1c demonstrates this concept using a series resistor-light emitting diode (LED) circuit on a polydimethylsiloxane (PDMS) substrate and a dispersion of 80 mg/ml of copper microspheres (radius 5 $\mu$m) in silicone oil (see Appendix). For demonstrations, the interconnect used was a single strand (100 $\mu$m) of a multi strand wire. Further, a well containing the dispersion was created over an exposed region of the interconnect and an open fault was deliberately introduced in this region. The entire sample was mechanically clamped to the translation stage for stretching. When an external voltage was applied, the LED did not initially light up due to the open fault. However, the electric field across the gap activated the self healing mechanism resulting in the LED eventually lighting up. When the entire system was stretched, the connectivity was maintained (with intermittent breaks) due to the heal increasing its length. The events occurring inside the well are also shown in Fig. 1c. 

Henceforth we discuss the mechanics of self healing and a key process of a self driven sintering. This sintering is important to achieve mechanical flexibility in the heal. We characterize the self healing mechanism and the response of the heal to stretching. All experiments in this work use a dispersion of copper microspheres in silicone oil (see Appendix).

\section{Mechanics of Healing with Sintering}
To study the mechanism of self healing, an open circuit fault was emulated using a test bed shown in Fig. 2a. A dispersion of copper microspheres (radius 5 $\mu$m) in silicone oil (see Appendix) was contained in a 200 $\mu$m wide gap between two electrodes. This gap mimicked the open fault while the electrodes represented the two ends of the disconnected interconnect. External resistors were placed in series to emulate the terminal impedances of the interconnect. The total external impedance was $R_{a}\sim$ 4.4 k$\Omega$. A Keithley 2410 source meter unit was used to source the voltage, $V_{a}$, and measure the dynamics of the current through the circuit throughout the experiment. Fig. 2b shows the typical observed dynamics of the current during the healing process for experiments performed with dispersions at different temperatures. The temperature influenced the viscosity of the fluid and therefore the dynamics of healing.

\begin{figure*}
\centering
\includegraphics[width=4 in]{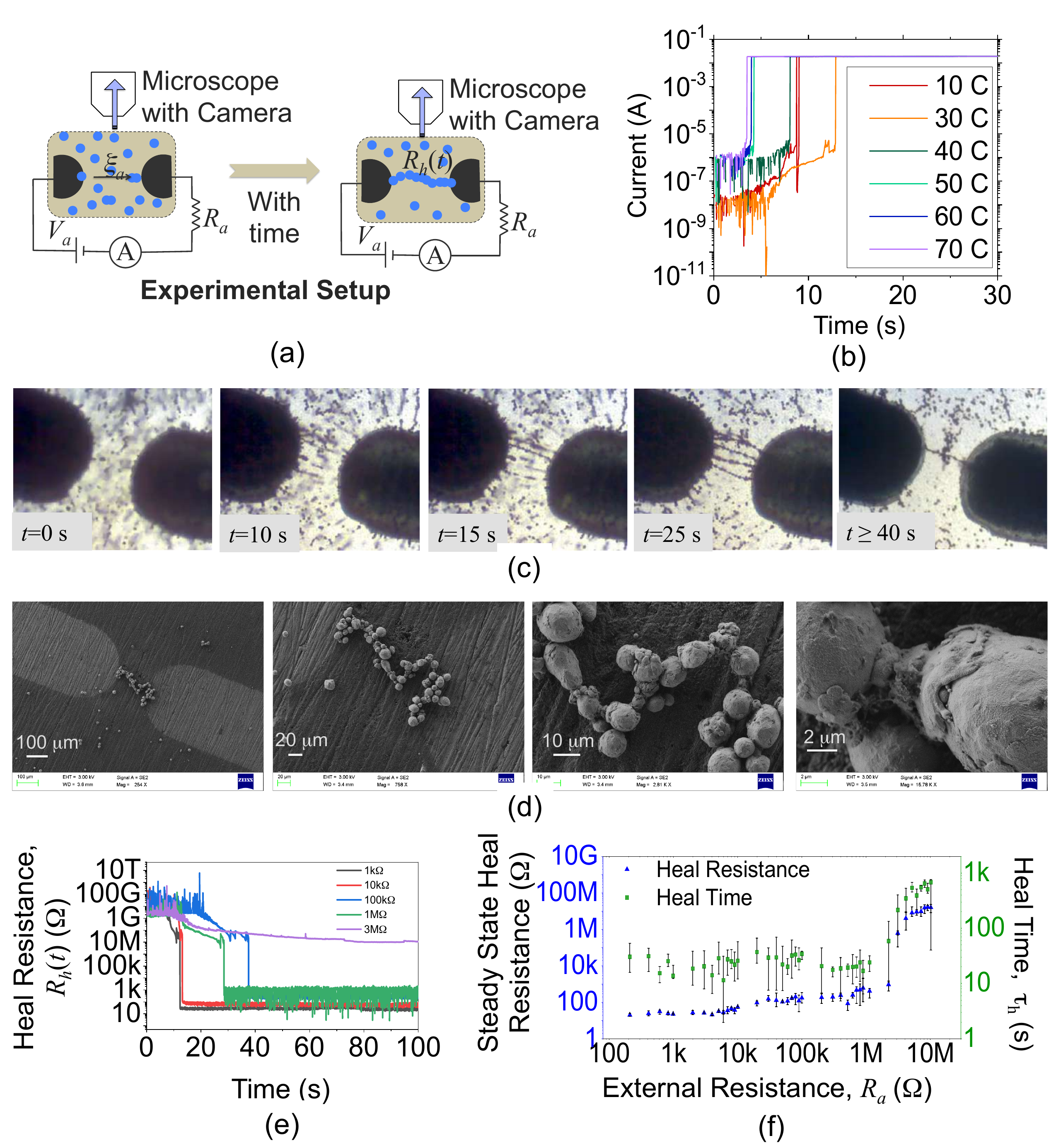}
\caption{(a) Experimental setup for the study of the mechanics of self healing. (b) Current versus time for several self healing experiments conducted with dispersions at different temperatures. At higher temperatures, the viscosity of the fluid is low and the healing is quicker. The sudden jump in current is indicative of the sintering of the heal. (c) Photographs of the process of self healing (d) SEM images of the sintered heal. (e) Transients of the heal resistance, $R_{h}(t)$, during healing for different external resistances, $R_{a}$. (f) The heal time and steady state heal resistance are dependent on $R_{a}$. Videos in Supplementary Material}
\end{figure*}

In real applications, the occurrence of an open fault in a current carrying interconnect would itself result in the electric field and initiate the healing process. However, for the purpose of this study, an electric field, $\xi_{a}$, was artificially created across the gap by applying an external voltage, $V_{a}$. The field strength was chosen with a view towards the possible application of self healing for printed electronic thin film transistor circuits on flex. In these systems, operating voltages are typically 5 V to 15 V and open faults causing 10 $\mu$m gaps (see Supplementary Material) would result in fields of approximately 0.5 V/$\mu$m to 1.5 V/$\mu$m. With this in mind, a field of $\xi_{a}\approx 0.4$ V/$\mu$m was created across the gap by applying an external voltage, $V_{a}=80$ V to the circuit of Fig. 2a. 

Upon experiencing this field the conductive particles of the dispersion in the gap were polarized. As shown in the sequence of photographs (Fig. 2c, Video in Supplementary Material), the induced dipole moment in the copper particles resulted in them chaining up due to dipole-dipole attraction and forming chains of loosely held linear clusters ($t$=25 s). Once these chains spanned the gap and bridged the two electrodes, the current through the interconnect was established to a small extent. This current caused joule heating. For certain conductive particles (eg. copper), this heating resulted in a weak sintering of the particles thereby leading to the reduction of the heal resistance with time. This sintered bridge constituted the heal. After sintering, the current through the interconnect increased rapidly leading to the restoration of electrical connectivity.  Fig. 2d shows SEM images of the sintered copper particles in the bridge (see Appendix). 

The total healing time, $\tau_{h}$, is the sum of two components - the time taken for the formation of the chains of electrostatically held particles and the time taken for sintering and formation of the single wire like bridge. The first component i.e. the time for the formation of chains is estimated by equating the the dipole-dipole attractive force that drives the mechanism of chain formation with the opposing force of viscous drag experienced by the particles moving through the fluid. This time can be shown to scale as $\sim (\eta_{f}/\epsilon_{f})(x_{p}/r_{p})^{5}\xi_{a}^{-2}$ with $\eta_{f}$ being the dynamic viscosity of the fluid, $\epsilon_{f}$ the permittivity of the fluid and $x_{p}$ the average distance between the conductive particles in a homogenous dispersion (See Supplementary Material). The second components, i.e., the dynamics of sintering is governed by a positive feedback mechanism that can be qualitatively described as follows. Upon the formation of the multiple chains of loosely held particles across the gap, the current through the interconnect distributes itself through these chains. However, minor differences in the resistance of each chain results in the current through the chains being different such that one of the chains, typically the shortest, carries the maximum current and heats the quickest. This results in quicker sintering and its resistance being lowered further which in turn increases the current and the heating through it. This results in the restoration and stabilization of the current through the interconnect. If the time dependent resistance of the heal is $R_{h}(t)$, the voltage drop across the gap would be $V_{a}R_{h}(t)/(R_{h}(t)+R_{a})$ and the rate of increase in temperature of the bridge would be proportional to $V_{a}^{2}R_{h}(t)/(R_{h}(t)+R_{a})^{2}$. This heating leads to the sintering and the reduction of $R_{h}(t)$ with time. The dependence of the dynamics of $R_{h}(t)$ on $R_{a}$ is shown in Fig. 2e. The dependence of the healing time and the steady state heal resistance on $R_{a}$ is shown in Fig. 2f. For successful self healing, the interconnect must have low terminal impedance i.e., it must be a current carrying interconnect.

\section{Self Healing with Stretching}
Fig. 3a shows the experimental setup to study the response of the heal to 1D stretching. To emulate the electrodes and the encapsulated dispersion as envisioned in Fig. 1a, a half-open micro channel (100 $\mu$m width and 500 $\mu$m depth) was molded on the surface of a PDMS substrate. The channel was then filled (from one side via capillary action) with the dispersion of copper microspheres in silicone oil. Two wires (100 $\mu$m diameter) were inserted into the channel from the two ends to form the two electrodes. One of the wires was attached to a precision translation stage (Thor Labs BSC 101 SIN 40833095) that enabled it to move back and forth in the channel with a precise velocity. The surface was finally capped with Kapton tape. The tape confined the dispersion and restricted the movement of the electrode to 1D. The external circuit was similar to Fig. 2a but with $R_{a}=220$ k$\Omega$.

To begin, the spacing between the electrodes was adjusted to $40$ $\mu$m (initial position) to emulate an open fault. Healing was first achieved at this initial position and a sintered bridge was allowed to form across the 40 $\mu$m gap (time $t$=0 s). When the current through the interconnect was stable, the electrodes were moved apart at a constant relative velocity of $u=5$ $\mu$m/s (after initial acceleration from rest). Fig. 3b illustrates a sequence of photographs from the experiment. The heal appeared to `stretch' in order to accommodate the movement of the electrodes (Video in Supplementary Material). The dynamics of the current through the circuit was continuously recorded throughout the experiment (Fig. 3c). The shaded subplot is the zoomed view of the time span from 50 s to 100 s. Despite intermittent breaks, the connectivity was repeatedly restored till around 900 s. Beyond this, the connectivity was permanently lost.

Following up from the discussion around Fig. 1b, this phenomenon of persistent restoration of electrical conductivity can be explained as follows. Initially, with the electrodes separated by an open gap, the particle chain bridges the gap, sinters and restores electrical connectivity. When the electrodes begin to move apart due to stretching, the sintered chain momentarily experiences large stress and breaks. However, a strong electric field is re-established in this new gap that is formed due to the break. The presence of the electric field once again initiates the self healing mechanism and involves locally available free particles to participate in the healing by polarizing them. The new particles that fill the gap soon sinter to become a permanent part of the heal (Fig. 3d). The heal therefore responds to stretching by adding more particles to its length.


\begin{figure*}
\centering
\includegraphics[width=7 in]{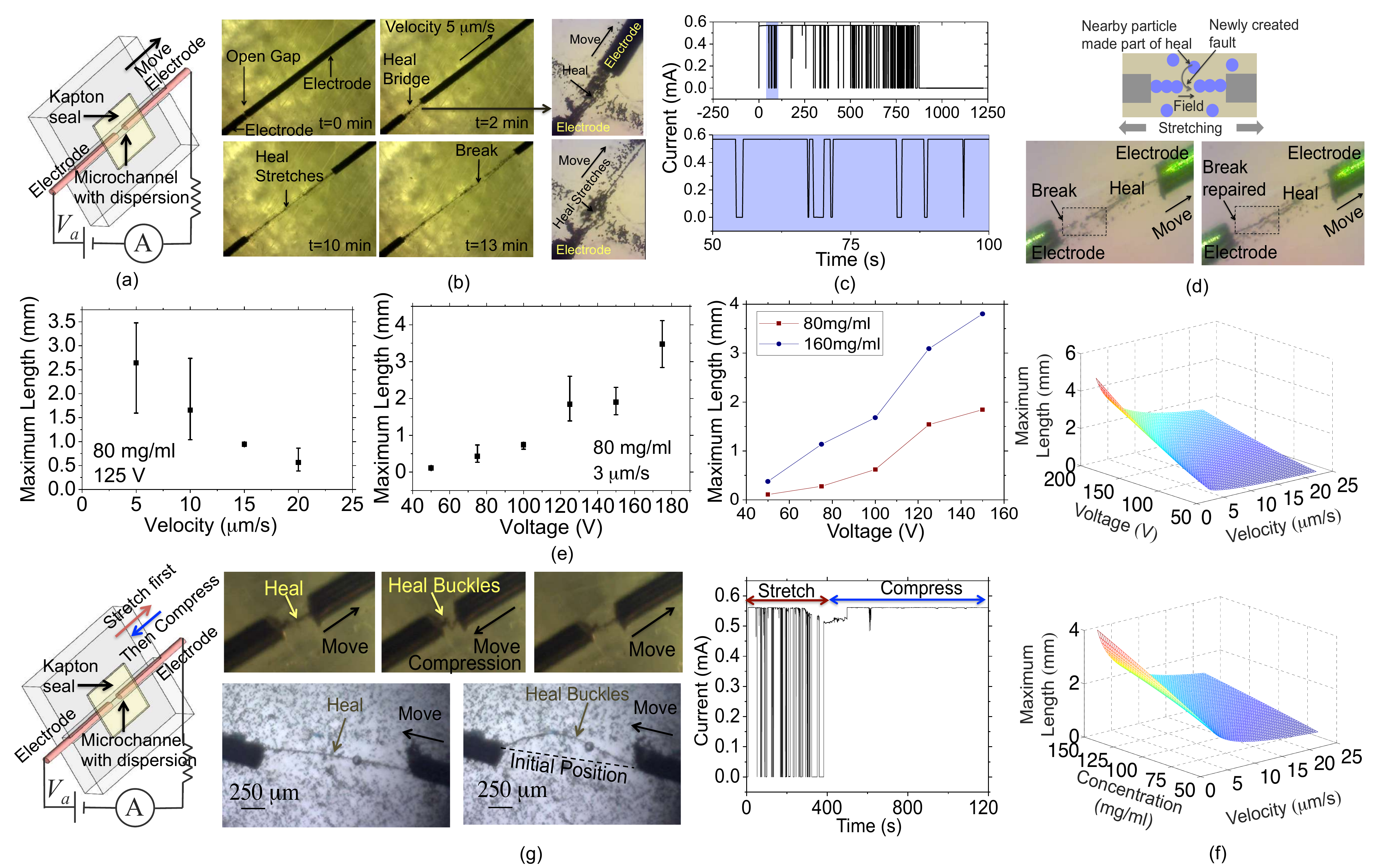}
\caption{(a) Experimental setup for the study of the impact of stretching on the heal. (b) Photographs from experiments demonstrating the stretchability of the heal. At $t$=0 min, the open gap of length 40 $\mu$m is healed. The electrodes are then moved apart at a constant velocity of 5 $\mu$m/s. (c) In response to stretching, the heal adaptively increases its length by involving more particles to participate in the healing process. (d) Measurements of current versus time during the stretching.  (e) Maximum stretchable length, $\Delta l_{max}$, as a function of the stretching velocity, $u$, applied voltage, $V_{a}$ and dispersion concentration (f) Prediction of $\Delta l_{max}$, as a function of $u$, $V_{a}$ and dispersion concentration based on Eq. 1. (g) Buckling of the heal in response to compression and the typical dynamics of the current through the heal in response to stretching and subsequent compression. Videos in Supplementary Material.}
\end{figure*}

The repeated drops and increases in current (Fig. 3c) also corroborates with this explanation. However, this repeated restoration of electrical conductivity does not last forever. If the local concentration of particles become too low or if the velocity of separation, $u$, is too high, electrical connectivity is permanently lost (here at around 900 s). In the former case, there are no particles available to fill in the gaps occurring in the expanding heal. In the latter case, the increasing distance between the electrodes rapidly weakens the field and weakens the dipole moment and the driving force for healing. Therefore, if the break is not filled in by a nearby particle immediately, it becomes increasingly harder for permanent restoration. The maximum length, $\Delta l_{max}$, to which the heal can stretch to without permanent loss of electrical connectivity depends on $V_{a}$, $u$, and the dispersion concentration. As shown in Fig. 3e, $\Delta l_{max}$ reduced with the increase in $u$, increased with an increase in $V_{a}$ and increased with an increase in dispersion concentration.

To quantify $\Delta l_{max}$, we consider the case where there exists a sintered heal across the open gap between two electrodes. At time $t=0$, the electrodes are moved apart with a constant velocity $u$. The stress in the sintered heal causing a fracture at some location in the heal and this new gap grows in length as $ut$ and the field across the gap reduces as $\sim V_{a}/ut$ (ignoring the non-uniformity of field). This field is responsible for re-activating self healing and filling in this gap with the locally available particles. However, if this field drops below a threshold electric field $\xi_{th}$, the dipole-dipole attractive forces become too weak to overcome the static friction and the Brownian motion and conductivity is not restored. The time taken for the electric field to drop to below $\xi_{th}$ scales as $\tau_{th}\sim V_{a}/(u\xi_{th})$. If the nearest free particles is at a distance of $\sim x_{p}$, the time constant for the particle to arrive at the gap can be shown to be $\sim (\eta_{f}/\epsilon_{f})(x_{p}/r_{p})^{5}u^{2}t^{2}/V_{a}$. If the time for sintering is ignored, this time constant defines the total repair time of the newly formed gap. Therefore for the heal to stretch successfully by the addition of more particles along its length, this repair time must be $\leq \tau_{th}$. Clearly as time $t$ increases, this condition is less likely to be satisfied. Therefore, the heal can stretch till equality is obtained in time constants and therefore,
\begin{equation}
\Delta l_{max}\propto (\epsilon_{f}/\eta_{f})^{1/2}(r_{p}/x_{p})^{5/2}(u\xi_{th})^{-1/2}V_{a}^{3/2}
\end{equation}
As an example, a dispersion of 80 mg/ml of $r_{p}=5$ $\mu$m particles silicone oil of kinematic viscosity 300 cSt implies $\eta_{f}$=0.28 kg/ms, $\epsilon_{f}$=$20\times10^{-12}$ F/m, $x_{p}$=38.8 $\mu$m. If $\xi_{th}$=0.04 V/$\mu$m. With the constant of proportionality in Eq. 1 being 0.06, the dependence of $\Delta l_{max}$ on $V_{a}$ and $u$ is shown in Fig. 3f. On the other hand, for a constant voltage of 125 V, the dependence of $\Delta l_{max}$ on the dispersion concentration (in weight/volume) and $u$ is shown in Fig. 3f.

\begin{figure*}
\centering
\includegraphics[width=4.5 in]{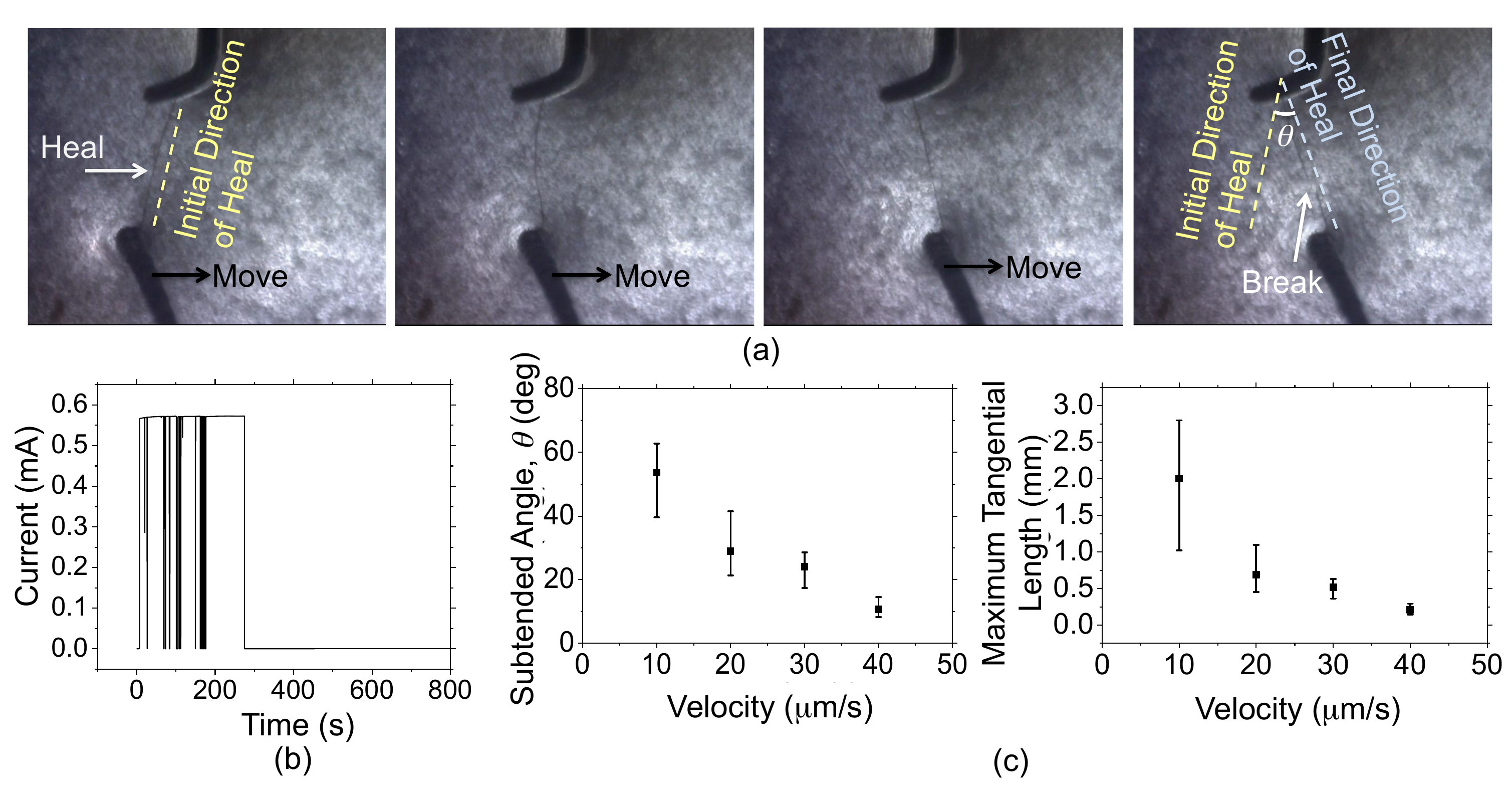}
\caption{(a) Demonstration of stretching in two dimensions. The electrodes are moved axially and then tangentially. The initial direction of the heal across the open gap is shown before the onset of tangential movement. The final direction of the heal is shown defining the moment the chain breaks and cannot stretch any further.  The maximum angle subtended between the initial and final direction is $\theta$. (b)  Plot of the typical current transients during the two dimensional movement. (c) The plot of the subtended angle $\theta$ as a function of the velocity and the maximum displacement length as a function of the velocity. Video in Supplementary Material.}
\end{figure*}

We next consider the case of 1D oscillations, i.e., 1D stretching followed by 1D compression of the heal as described by Fig. 3g. After being stretched and subsequently compressed, the heal grew in length to accommodate the stretching and subsequently buckled to accommodate compression, all the while maintaining electrical connectivity. Fig. 3g shows the buckling in the heal during compression observed during different experiments. The typical dynamics of the current through the heal in response to stretching and subsequent compression is also shown. 

For experiments on 2-D stretching, a 5 mm by 5 mm by 2mm (depth) well in PDMS was used to contain the dispersion. 100 $\mu$m single strand wires were slightly bent at the ends and immersed in the dispersion to form the electrodes. One of the wires was connected to two approximately orthogonally aligned translation stages (Axial control: Holmarc MTS 3760; Tangential control: Thor Labs BSC 101 SIN 40833095). The external circuit was similar to that used in Fig. 3a. At the start of the experiment, the electrodes were placed 40 $\mu$m apart and the voltage was applied resulting in a heal. Once the current stabilized, the electrodes were moved apart axially allowing the heal to stretch. After a 1.15 mm displacement, axial translation was stopped and tangential translation was activated. The images of Fig. 4a illustrate the impact of tangential movement. The heal therefore experienced two kinds of forces. The first was the stress due to axial stretching. The second was the distributed loading along its length due to viscous drag experienced by the heal as it moves through the fluid during tangential movement.Except for intermittent loss of conductivity, this two dimensional movement did not disrupt the electrical connectivity between the electrodes. However, for large displacements, the heal was permanently fractured resulting in a loss of conductivity. Fig. 4b plots the current as a function of time during the experiment. Fig. 4c plots the maximum subtended angle between the initial position of the heal to the final position (when the heal fractures permanently) as a function of the velocity of movement along the tangential direction. Fig. 4c also plots the maximum displacement along the tangential direction before failure. 

\section{Discussion}
This work presented the possibility of realizing interconnects with the ability to automatically heal an open circuit failure with the heal having an ability to stretch. This was achieved by the use of a rather simple system - a dispersions of conductive particles in an insulating fluid. If an open fault were to occur in a current carrying interconnect, the resulting electric field in the gap would initiate the self healing mechanism by polarizing the conductive particles located in the gap and resulting in them chaining up (dipole-dipole attraction) to eventually form a bridge across the gap. The repair mechanism is therefore truly self driven, i.e. it self-activates and self-terminates. The occurrence of the fault creates the electric field that initiates the self healing mechanism. The healing mechanism restores conductivity by forming a sintered chain of particles (the heal) across the open gap. Once the gap is short circuited by the heal, the field disappears discouraging further chain formation and stopping the mechanism. The healing mechanism is akin to blood clotting where the various species in the blood rush to the site of injury to participate in clotting.

The key feature of the healing mechanism is the possibility of compressing and stretching the heal without losing electrical connectivity. A requirement to achieve this was the sintering of the particles constituting the bridge. This too was self driven due to Joule heating governed by positive feedback for as long as $R_{h}(t)>R_{a}$. Sintering made the heal behave like a conductive thread or chain. In compression, this chain could accommodate compression by buckling appropriately. In tension (stretching), the chain would break. Since the chain segments on either side of this break were conductive, the electric field would be re-established in this newly formed gap thus reactivating the healing by encouraging locally available particles to fill this gap and complete the chain. Thus the sintered chain would accommodate stretching by effectively adding more particles to its length. This is a fundamental difference compared to using conductive materials that stretch due to low modulus of elasticity.  

For the sake of comparison with other stretchable conductors, it is useful to model the effective mechanical properties of the heal.
Using 80 mg/ml dispersions of copper microspheres in silicone oil, for $V_{a}=125 V$, the maximum displacement ranged from 0.5 mm (for $u=20$ $\mu$m/s) to 2.5 mm (for $u=5$ $\mu$m/s). Since the initial separation of the electrodes was 40 $\mu$m, this is equivalent to a minimum strain of 12.5 to a maximum strain of 60.25 depending on the strain rate. Although there is an influence of the strain rate, the stress dynamics due to a step in strain is not like a viscoelastic material. Instead, the averaged effective behavior of the heal is to respond to the step in strain by adding more particles to increase its length and with no increase in average stress. Therefore it is more apt to model this behavior as plastic with the rupture point dependent on the strain rate. 

For complete integration of the self healing mechanism with integrated circuits on flex, three major problems need to be overcome. The first is with regards to the encapsulation and techniques of packaging of the dispersion with high density interconnects in the manner shown in Fig. 1a. The solution to this problem may lie in borrowing ideas from inkjet printed electronics and microfluidics. These technologies are compatible with flexible electronics and would permit the controlled dispensing and packaging of the dispersion. The second problem is with regards to scaling down. All experiments described in this paper used interconnects and open gaps of width of $\sim 100$ $\mu$m. This permitted the use of dispersions having metallic particles of diameter of $\sim 10$ $\mu$m. However, thin film transistor based integrated circuits on flex typically use interconnects of width $\sim 100$ $\mu$m which would results in open faults of widths of the same length scale. For this study to scale down, the dispersion must now use particles of $\sim 1$ $\mu$m diameter. This poses problems with the stability of the dispersion as well as impacting the time taken to heal due to a much weaker dipole-dipole forces and a more influential Brownian motion. This problem can be overcome by increasing the dispersion concentration. However, such an approach spills over to the third major problem which is with regards to cross-talk. As the dispersion concentration is increased and as the dispersion heads towards becoming a conductive fluid, the significantly three dimensional feature of the interconnects due to the encapsulation would result in a significant cross talk between two adjacent interconnects thereby limiting the frequency response. Therefore, the use of a very high dispersion concentration is not ideal and this parameter needs to be optimized.

Nevertheless, this approach to self healing is versatile and has the potential for application across a varying class of systems such as flexible and wearable electronics to commercial printed circuit boards. This versatility stems from two features. First, the approach does not use rare materials (eg. Ga-In, graphene etc) but instead a dispersion of copper particles in a common insulating fluid (transformer oil). Second, the approach does not change the conventional material and processes used for interconnect fabrication. Instead, the concept can be implemented as an add on feature require additional fabrication processes if self healing is desired. These features permit interconnects with the ability to self heal and stretch on flexible substrates thereby improving interconnect reliability significantly.

\section{Conclusion}
This work demonstrated the possibility of self healing interconnects with the ability to stretch significantly. This was demonstrated using copper-silicone oil dispersions. Heals having near metallic conductivity of $\sim 5 \times 10^{5}$ Sm and a stretchability with strains from 12 to 60 depending on the strain rate were demonstrated.  Previously, stretchable interconnects used materials other than copper. Here we effectively show self healing, stretchable copper. This work promises high speed, self healing and stretchable interconnects on flex thereby improving system reliability.

\section*{Acknowledgement}
This work was funded by the EPSRC Grant No. RG92121 and DST IMPRINT Grant No. 7969. Amit Kumar and Virendra Parab contributed equally to this work. Sanjiv Sambandan thanks the DBT India-Cambridge Lectureship program for permitting a joint appointment between the Indian Institute of Science and the University of Cambridge.

\section*{Appendix: Materials and Methods}
\subsection*{Dispersion Preparation}
Dispersions of copper microspheres (radius 5 $\mu$m, Alfa Aesar 042689) in silicone oil (kinematic viscosity 300 cSt, S. D. Fine-Chem Limited 25072) were used. Desired concentrations (in mg/ml) were obtained by measuring the weight of copper particles to be dispersed in a known volume of silicone oil. The dispersion was sonicated for 1 hour at 80 C followed by mechanical stirring to achieve the required homogeneity. At room temperature, it was observed that the viscosity of the fluid aided the preparation of relatively stable and homogenous dispersions that were sufficiently suitable for the applications discussed in this work. No other stabilization techniques were used. 

\subsection*{Sample Preparation for SEM Imaging}
For SEM images (Fig. 2d), the experiment was first performed as described in Fig. 2 on a printed circuit board.  The current versus time was constantly measured. 30 s after  the current was restored and stabilized (indicative of sintering), the circuit was disconnected. The regions between the electrodes now contained the sintered heal immersed in silicone oil. To obtain SEM images this region was repeatedly rinsed with isopropyl alcohol to remove the oil. The sample was then dried and the section of the printed circuit board containing the bridge was transferred to an SEM stub for imaging. A conductive carbon tape was used to short circuit the electrodes (and therefore the heal) to the SEM stub for good imaging.

\bibliography{ref}

\end{document}